\documentclass[%
 reprint,
superscriptaddress,
 amsmath,amssymb,
 aps,
]{revtex4-2}

\usepackage{graphicx}% Include figure files
\usepackage{dcolumn}% Align table columns on decimal point
\usepackage{bm}% bold math
\usepackage{physics}
\usepackage[colorlinks=true,
            linkcolor=blue,
            citecolor=red,
            urlcolor=blue]{hyperref}

\newcommand{\ITMO}{School of Physics and Engineering, ITMO University,
197101 St. Petersburg, Russia} 

\newcommand{\Tongji}{Institute of Acoustics, School of Physics Science and Engineering, Tongji University, Shanghai 200092, China}
\newcommand{\iu}{{i}\mkern1mu}
\newcommand{\eu}{\mathrm{e}\mkern1mu}

\begin{document}

%\definecolor{mybluegray}{rgb}{.58,.67,.81}

\preprint{AIP/123-QED}

\title[Unidirectional Transverse Scattering in Acoustic Dimers]{Unidirectional Transverse Scattering in Acoustic Dimers}
% Force line breaks with \\
\author{Mikhail Smagin}
\affiliation{\ITMO}

\author{Iuliia Timankova}
\affiliation{\ITMO}

\author{Pavel Pankin}
\affiliation{\ITMO}

\author{Yong Li}
\affiliation{\Tongji}
\affiliation{\ITMO}

\author{Mihail Petrov}
\affiliation{\ITMO}%
 \email{m.petrov@metalab.ifmo.ru}

%\date{\today}% It is always \today, today,
             %  but any date may be explicitly specified

\begin{abstract}
We study unidirectional transverse scattering in a two-dimensional acoustic dimer composed of two circular subwavelength scatterers. Using a coupled multipole model, we show that interparticle coupling enables effective monopole--dipole interference and supports a transverse Kerker effect under plane-wave excitation. In contrast to a single non-absorbing rotationally symmetric particle, where Kerker-type directional scattering is only approached in the weak-scattering limit, the dimer can combine pronounced directionality with strong overall scattering. This regime is promising for compact acoustic beam steering and directional wave routing.

\end{abstract}

\maketitle

Directional scattering is a key concept behind wave manipulation~\cite{liu2018}, with applications such as beam steering~\cite{liu2017beam}, frequency-~\cite{hancu2014multipolar} and
polarization~\cite{barreda2017electromagnetic}  selective transport, and directional emission~\cite{rutckaia2017quantum}. At the subwavelength scale, directional scattering is achieved through interference of a few
resonant multipoles, which shape the angular distribution of the scattered field. In optics, the simplest case was described by Kerker \textit{et al.} in 1983~\cite{kerker1983}, showing that a combination of electric and
magnetic dipoles can form a cardioid scattering pattern when phase and amplitude conditions are satisfied. This concept was later generalized to combinations of higher-order multipoles~\cite{liu2018}.

The first and second Kerker conditions as originally formulated correspond to cancellation of backward and forward scattering, respectively. The idea of the Kerker effect was then analytically
extended~\cite{bag2018transverse,lee2018simultaneously}, and experimentally demonstrated~\cite{shamkhi2019transverse,bukharin2022transverse,matsumori2023unidirectional}, to directional scattering in the transverse
direction. This \emph{transverse Kerker effect} corresponds to scattering suppression in both forward and backward directions simultaneously~\cite{shamkhi2019transverse,lee2018simultaneously}. Such an electromagnetic
transverse Kerker effect can be either bidirectional, corresponding to dumbbell radiation pattern~\cite{shamkhi2019transverse,lee2018simultaneously,bukharin2022transverse,liu2019lattice}, or unidirectional, corresponding
to transverse cardioid radiation pattern~\cite{bag2018transverse,zheng2023annular,du2024manipulation,matsumori2023unidirectional}. Both types of the effect have been found to be useful in a wide range of applications, including
lattice invisibility~\cite{liu2019lattice,shamkhi2019transparency,zhang2021wide}, directional wave routing~\cite{nechayev2019huygens,yu2022broadband,li2016all}, antennas for directional emission~\cite{qin2022transverse},
precise displacement  sensing~\cite{bag2018transverse,zhang2024magnetic,zhang2021nanometric,shang2019unidirectional}, and lateral optical forces~\cite{matsumori2023unidirectional,zheng2023annular}.

In acoustics, classical forward and backward Kerker directionality was introduced only recently~\cite{wei2020,wu2021}. In this case directional scattering can similarly be understood through multipole interference, which in the subwavelength regime is dominated by monopole and dipole responses~\cite{blackstock2000,toftul2019}. Multipole interference has also been shown to induce directional acoustic forces~\cite{Smagin2024,toftul2025acoustic,toftul2026radiation} and to produce BICs in lattices of acoustic resonators~\cite{ustimenko2026lattice}. Directional forward and backward scattering has been experimentally demonstrated recently as well~\cite{timankova2026experimental}. The directionality of acoustic scattering in transverse directions, however, remains underreported.  

Here, we show that a simple acoustic dimer provides conditions for a unidirectional transverse acoustic Kerker effect. We start off by stating that in non-absorbing \emph{rotationally symmetric} acoustic scatterers, i.e., 3D spheres and 2D infinitely long cylinders, even classical \emph{longitudinal} Kerker conditions are only satisfied in the limit of negligible scattering, in strong contrast to optics. However, a dimer structure can combine directional scattering in forward and backward directions while maintaining strong overall scattering. We then predict that coupling between two circular subwavelength scatterers creates effective monopole--dipole interference that supports transverse Kerker scattering. The transverse Kerker effect is then showcased numerically on a dimer of two realistic labyrinthine particles. These results establish acoustic dimers as a promising platform for the study of the acoustic Kerker effect and potentially open a route for novel compact beam-steering devices.

% --- Fig. 1: concept/geometry ---
\begin{figure}[t!]
  \centering
  \includegraphics[width=0.85\linewidth]{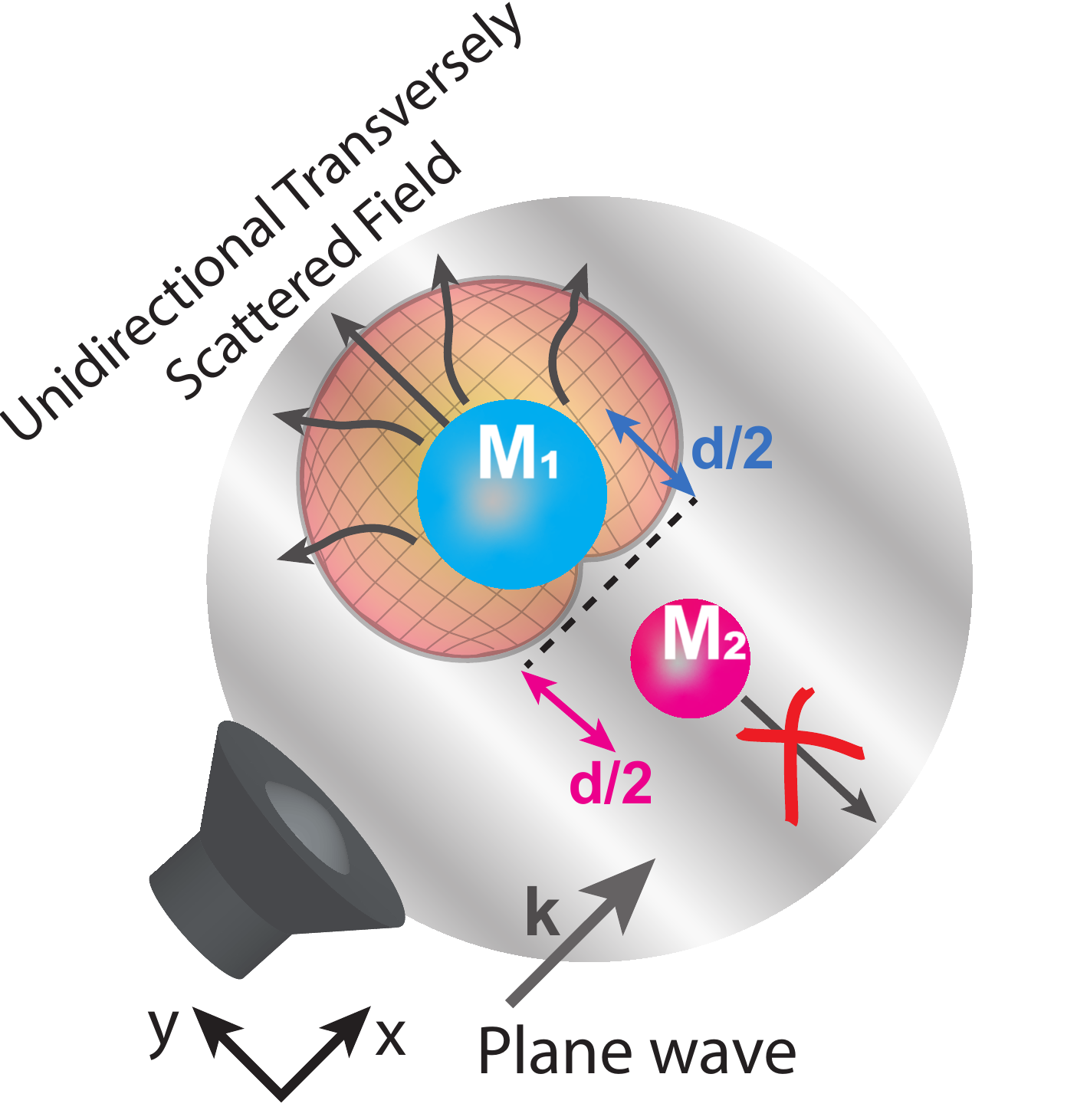}
  \caption{\textbf{Geometry of the problem.} Acoustic dimer separated by $d$ and illuminated by a plane-wave. For transverse incidence ($\mathbf{k}\parallel\hat{\mathbf{x}}$), the dimer can radiate unidirectionally along $\pm\hat{\mathbf{y}}$ due to the induced effective dipole moment along the dimer axis $\hat{\mathbf{y}}$.}
  \label{fig:concept}
\end{figure}

% --- (2) Single-particle Kerker statement kept as Eq.(1) ---

% --- Fig. 2: isotropic limitation ---
\begin{figure*}[t!]
  \centering
  \includegraphics[width=1.0\linewidth]{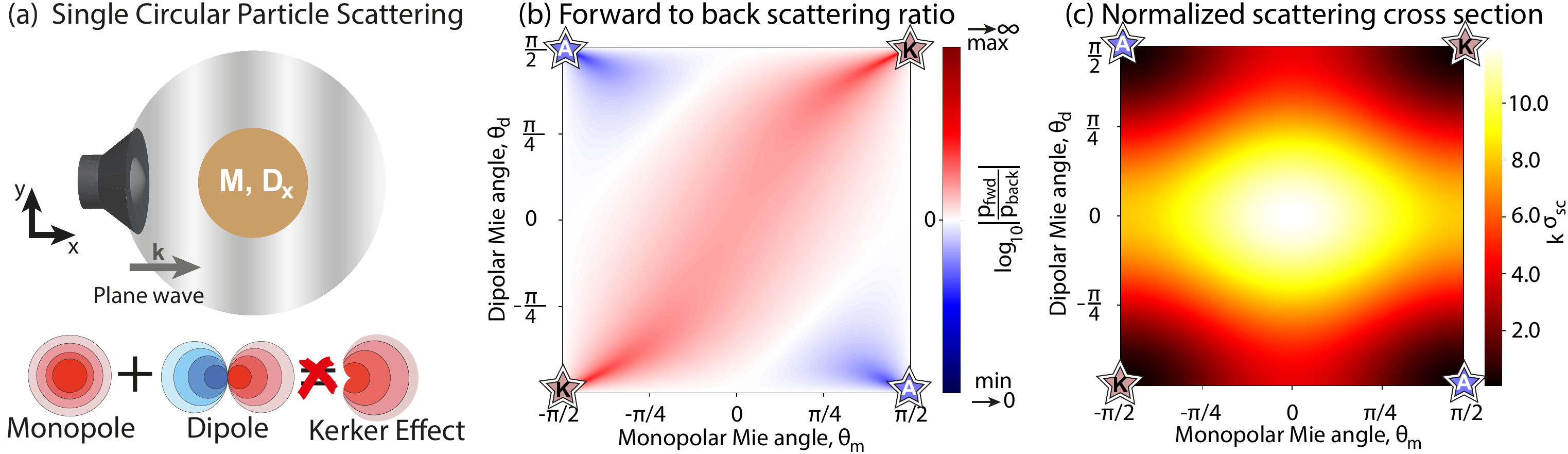}
  \caption{\textbf{Non-absorbing rotationally symmetric particle limitation.}
(a) Single non-absorbing rotationally symmetric particle illuminated by a plane-wave does not exhibit perfect directional scattering.
(b) The logarithmic ratio of forward to backward scattering on a non-absorbing 2D rotationally symmetric particle in the parameter space of Mie angles. Strong directional scattering is exhibited in the corners far from monopole and dipole resonances.
(c) The normalized total scattering cross section $k\,\sigma_{\mathrm{sc}}$ of a non-absorbing rotationally symmetric 2D particle in the parameter space of Mie angles. The scattering cross section is inherently small in the strong-directionality region, i.e., approaching the Kerker condition forces the scattering amplitude to vanish. The star-shaped points correspond to the first ($\mathbf{K}$) and second Kerker conditions (anti-Kerker, $\mathbf{A}$).}
  \label{fig:isotropic}
\end{figure*}

We start by postulating that for a \emph{non-absorbing rotationally symmetric} acoustic particle under plane-wave excitation, the monopole--dipole far-field scattering cancellation conditions in the forward and backward directions are never satisfied exactly, so that the Kerker effect cannot be perfectly achieved; see Fig.~\ref{fig:isotropic}(a) and Appendix~\ref{appendix:proof} for more details. In this section, we consider the case of two-dimensional rotationally symmetric particles (infinitely long 2D cylinders); however, this result is valid for three-dimensional particles (spheres) as well. For a plane-wave incident along $+x$, these conditions read:
\begin{equation}
p_{\mathrm{fwd}} \propto M + \frac{D_x}{c} = 0,
\qquad
p_{\mathrm{back}} \propto M - \frac{D_x}{c} = 0,
\label{eq:kerker_single}
\end{equation}
where $M$ is the monopole moment of the particle, and $D_x$ is the $\hat{\mathbf{x}}$-component of the dipole moment. Thus $p_{\mathrm{back}}=0$ corresponds to the first, and $p_{\mathrm{fwd}}=0$ corresponds to the second Kerker condition, or anti-Kerker condition. 

For a non-absorbing rotationally symmetric particle, it is convenient to introduce Mie angles $\theta_n\in(-\pi/2,\pi/2)$ through acoustic Mie coefficients
$a_n=-\cos(\theta_n)e^{i\theta_n}$~\cite{blackstock2000,toftul2019,yosioka}, which reduces the description of each scattering channel to just one parameter $\theta_n$~\cite{toftul2025acoustic,rahimzadegan2020minimalist}. The resonance in such a parametrization corresponds to $\theta_n=0$, and at $\theta_n=\pm\pi/2$ the scattering amplitude goes to zero. In the monopole--dipole approximation, the description of acoustic scattering on a rotationally symmetric particle reduces to just two parameters, namely the monopolar Mie angle $\theta_m$ and dipolar Mie angle $\theta_d$. Using this parametrization, we show that for a non-absorbing rotationally symmetric particle, the Kerker conditions are satisfied only in the limit where $(\theta_m,\theta_d)$ approaches one of the four corner points $(\pm\pi/2,\pm\pi/2)$, which corresponds to the limit of zero scattering. We illustrate this for a two-dimensional case by plotting a logarithmic ratio of forward to backward pressure scattered by an infinitely long 2D cylinder illuminated by a plane-wave; see Fig.~\ref{fig:isotropic}(b). The region close to the first and second Kerker conditions is confined to the corners of the $(\theta_m,\theta_d)$ plane, which is necessarily accompanied by intrinsically weak scattering; see the normalized scattering cross section $k\,\sigma_{\mathrm{sc}}$ in Fig.~\ref{fig:isotropic}(c). Here and further in this paper, we assume the wavenumber $k=2\pi$, such that all distances are expressed in $\lambda$. Without loss of generality, we assume air as the host material, with speed of sound $c=343$~m/s and material density $\rho_0=1.23$~kg/m$^3$.

To overcome the restrictions of non-absorbing rotationally symmetric particles, we consider a simple case of a 2D anisotropic particle---a \emph{dimer} of two circular subwavelength 2D particles (infinitely long cylinders) separated by distance $d$ along $\hat{\mathbf{y}}$, as shown in Fig.~\ref{fig:concept}. Due to their subwavelength nature, each particle can be modeled by monopole and in-plane dipole moments characterized by Mie angles $\theta_{m1(2)}$, $\theta_{d1(2)}$, and the induced moments $M_{1(2)}$, $\vb{D}_{1(2)}$, where subscripts $1(2)$ correspond to the top (bottom) particle. The induced moments are obtained self-consistently by solving the coupled multiple-scattering problem through the local field, which is the sum of the incident field and the field scattered by the other particle in the dimer. The moments induced in each particle are then translated to the center of the dimer to calculate the effective response of the dimer, $\bar{M}$ and $\bar{\vb{D}}$. All details of the analytical calculations and the general theory of a dimer with monopole and dipole moments are given in the Supplemental Information (SI).

\begin{figure*}[t!]
  \centering
  \includegraphics[width= 1.0\linewidth]{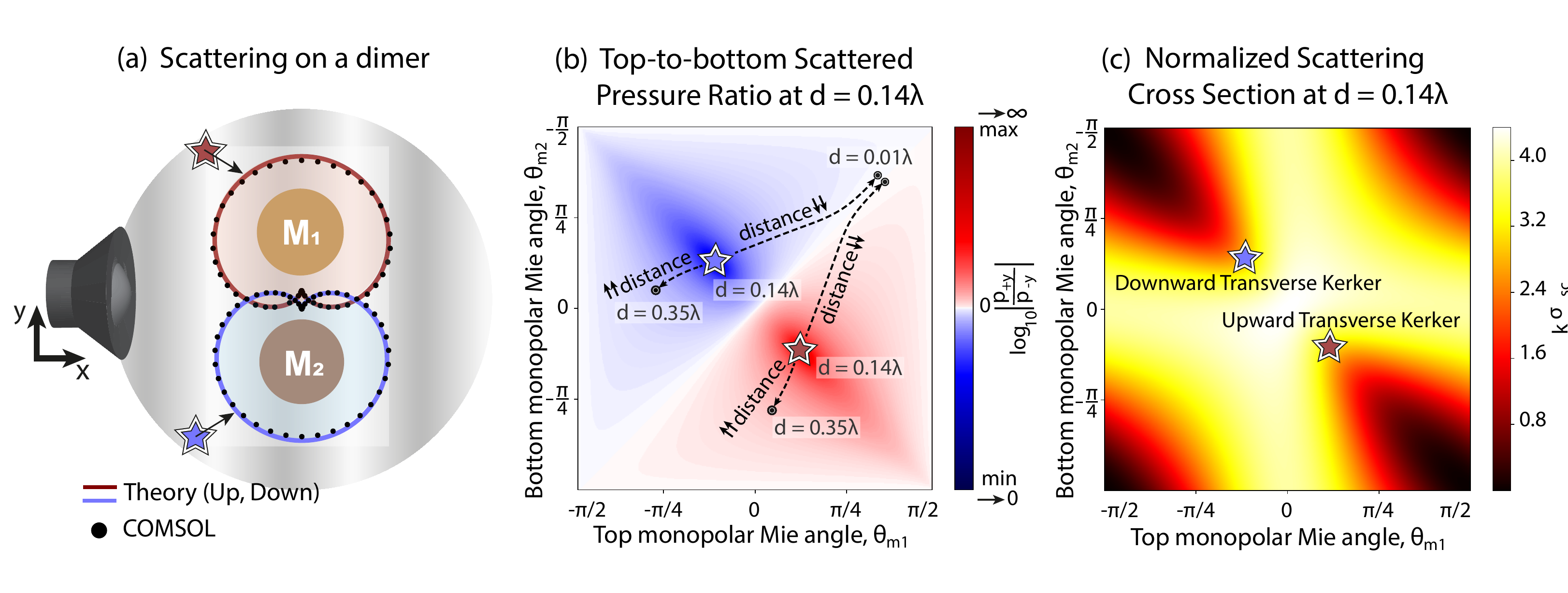}
  \caption{\textbf{Transverse Kerker Effect in Acoustic Dimers.}
(a) Acoustic monopolar dimer with interparticle distance $d$. The top (bottom) particle has monopole moment $M_1$ ($M_2$), and the intrinsic scattering properties of both particles are characterized by Mie angles $\theta_{m1}$ ($\theta_{m2}$). The incident wave is lateral to the dimer axis $\hat{\mathbf{y}}$. The analytically and numerically calculated radiation patterns are given for a dimer with interparticle distance $d=0.14\lambda$ at the upward (downward) Kerker conditions. The analytical calculations are performed for two circular 2D particles with $\theta_{m1(2)}=0.396$, $\theta_{m2(1)}=-0.375$, and the full-wave simulations are performed with infinitely long 2D cylinders with material and size parameters $r_1=0.0282\lambda$, $\bar{\rho}_1=0.837$, $\bar{\kappa}_1=25$, $r_2=0.038\lambda$, $\bar{\rho}_2=0.66$, and $\bar{\kappa}_2=35$; see the Supplemental Information for details.
(b) Top-to-bottom logarithmic scattered-pressure ratio of a two-dimensional dimer with interparticle distance $d=0.14\lambda$ in the parameter space of the top and bottom monopolar Mie angles under lateral plane-wave excitation.
(c) Normalized scattering cross section $k\,\sigma_{\mathrm{sc}}$ of a two-dimensional monopolar dimer with interparticle distance $d=0.14\lambda$ in the parameter space of monopole Mie angles under lateral plane-wave excitation. The stars correspond to the upward and downward Kerker conditions.}
  \label{fig:kerker_maps}
\end{figure*}

We limit our consideration to the simplest case of a dimer of two purely monopolar particles, such that $\theta_{d1}\simeq\theta_{d2}\simeq\pm\pi/2$, i.e., particles possessing only induced monopole moments $M_1$ and $M_2$. In this case, the effective multipole moments have simple expressions:
\begin{equation}
\bar M = (M_1+M_2)\,J_0\!\left(\tfrac{kd}{2}\right),
\quad
\bar D_y = 2ic\,(M_1-M_2)\,J_1\!\left(\tfrac{kd}{2}\right),
\label{eq:eff_moments}
\end{equation}
where $J_n$ are the Bessel functions.

Equation~\eqref{eq:eff_moments} shows that even a dimer of two purely monopolar particles in general still exhibits an effective dipole moment along the dimer axis $\hat{\mathbf{y}}$. If effective monopole and dipole responses $\bar{M}, \bar{D_y}$ satisfy conditions analogous to Eqs \eqref{eq:kerker_single}, the effective scattering
cancellation in one of the two opposite $\pm\hat{\mathbf{y}}$ directions can be achieved. Substituting Eq \eqref{eq:eff_moments} into Eqs \eqref{eq:kerker_single}, closed form conditions on the complex monopole ratio $M_1/M_2$ can be obtained. This gives:
\begin{equation}
\begin{aligned}
p_{\pm y}=0
\qquad &\Longleftrightarrow\qquad
\frac{M_1}{M_2}
=
-\,\frac{ J_0\!\left(\tfrac{k d}{2}\right)\mp 2 i\,J_1\!\left(\tfrac{k d}{2}\right)}
{ J_0\!\left(\tfrac{k d}{2}\right)\pm 2 i\,J_1\!\left(\tfrac{k d}{2}\right)},
\\ \text{which for real } kd &\text{ can be expressed as:} \\
|M_1| &= |M_2|,\\
\arg(M_1)-\arg(M_2)
&=
\pi + 2\,\atan2\!\left(\pm\,2 J_1\!\left(\tfrac{k d}{2}\right),\, J_0\!\left(\tfrac{k d}{2}\right)\right).
%\\\quad (\mathrm{mod}\;2\pi).
\end{aligned}
\label{eq:kerker_phase_mod}
\end{equation}

In the subwavelength limit $kd\ll 1$ the phase conditions reduce to $\arg(M_1)-\arg(M_2)\simeq \pi \pm kd$. This is the condition widely used in acoustic engineering to construct devices with cardioid directionality using two monopole microphones/speakers, by taking equal magnitude monopoles of reverse polarity with a phase delay $T = d/c$ \cite{benesty2012study, elko1995simple, kim2013sound, olson1973gradient}.
%, or a $\pi/4$ delayed monopole at $\lambda/4$ separation as a particular example[].
Eq.~\eqref{eq:kerker_phase_mod} gives the generalization of this condition beyond the small $kd$ limit in two dimensions.

To analyze scattering in acoustic dimer, we compute the induced monopole moments $M_1,M_2$ and search for solutions of Eq.~\eqref{eq:kerker_phase_mod} in the monopole-angle plane $(\theta_{m1},\theta_{m2})$.
One can also characterize directionality by the ratio of scattered pressure in the $\hat{\mathbf{y}}$ direction, i.e. $p_{+y}/p_{-y}$:
\begin{equation}
\frac{p_{+y}}{p_{-y}}
=\frac{J_0\!\left(\tfrac{kd}{2}\right)(M_1+M_2)+2iJ_1\!\left(\tfrac{kd}{2}\right)(M_2-M_1)}
     {J_0\!\left(\tfrac{kd}{2}\right)(M_1+M_2)-2iJ_1\!\left(\tfrac{kd}{2}\right)(M_2-M_1)}.
\label{eq:top_bottom_ratio}
\end{equation}
The minima and maxima of Eq.~\eqref{eq:top_bottom_ratio} may then point to the parameters where pressure scattered in one of the directions $\pm \hat{\mathbf{y}}$ is cancelled, and thus conditions \eqref{eq:kerker_phase_mod} are satisfied.

%The resulting design maps are summarized in Fig.~\ref{fig:kerker_maps}.
If we consider the \textbf{longitudinal incidence} along dimer axis, $\mathbf{k}\parallel\hat{\mathbf{y}}$, the dimer supports full cancellation in the backward direction -$\hat{\mathbf{y}}$, while retaining strong scattering for a wide range of parameters, see Appendix~\ref{appendix:longitudinal}.

This behavior contrasts sharply with the single-particle limitation, where approaching cancellation necessarily forces the scattering amplitude to be zero, see Fig.~\ref{fig:isotropic}.

\begin{figure*}[t!]
  \centering
  \includegraphics[width=0.83\linewidth]{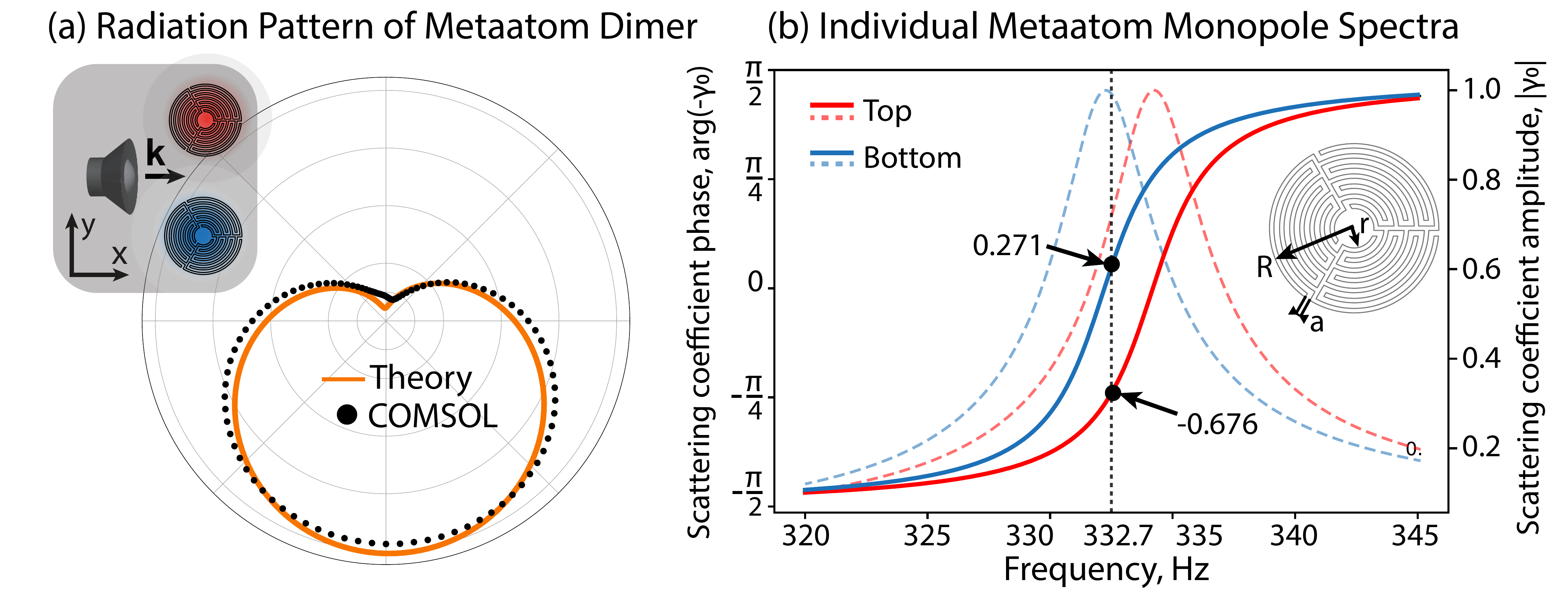}
  \caption{\textbf{Numerical validation of the transverse Kerker effect using two-dimensional labyrinthine meta-atoms.}
(a) Far-field radiation pattern of a dimer of two monopolar labyrinthine meta-atoms of different sizes; see the inset. The analytical calculation is performed in the circular monopolar-particle model, and the numerical calculation is performed for a solid labyrinthine-particle dimer in COMSOL. The dimer has interparticle distance $d=0.2\lambda$.
(b) Phase and amplitude spectra of the monopolar scattering coefficient $\gamma_0$ for the individual top and bottom meta-atoms illuminated by a plane-wave incident along $\hat{\mathbf{x}}$. The bottom meta-atom has channel width $a=0.002$~m, inner radius $r=0.007$~m, outer radius $R=0.03$~m, three sectors, and a curling number, i.e., the number of full channel turns, of 6. The top meta-atom has the same design as the bottom one, but its size is scaled by $0.994$.}
  \label{fig:metaatom}
\end{figure*}

More interesting is the case of \textbf{lateral excitation}, $\mathbf{k}\parallel\hat{\mathbf{x}}$; see Fig.~\ref{fig:kerker_maps}(a). In this case, the transverse Kerker effect can be achieved, which is not supported for a rotationally symmetric particle. For the interparticle distance $d=0.14\lambda$, we plot the absolute value of the logarithmic top-to-bottom scattering ratio, Eq.~\eqref{eq:top_bottom_ratio}, in Fig.~\ref{fig:kerker_maps}(b). Under such excitation, the clear minimum and maximum correspond to downward and upward \textbf{transverse} unidirectional Kerker scattering. Radiation patterns for the downward and upward Kerker conditions are shown in Fig.~\ref{fig:kerker_maps}(a). We also prove that these extrema in fact correspond to true zeros of $p_{\pm y}$, i.e., perfect Kerker scattering cancellation; see the Supplemental Information for details. For $d=0.14\lambda$, the downward (upward) Kerker effect is satisfied at $\theta_{m1(2)}=0.396$ and $\theta_{m2(1)}=-0.375$, which corresponds to strong resonant scattering, as can be seen from the colormap of the normalized scattering cross section in Fig.~\ref{fig:kerker_maps}(c), in contrast to the non-absorbing rotationally symmetric particle case. This constitutes the central result of our paper: the prediction of the \emph{transverse Kerker effect} in a dimer structure.

 To validate our results, we optimize the size and material parameters of infinitely long 2D cylinders for the aforementioned Mie angles at the Kerker condition. We characterize the particles by the relative material parameters $\bar{\rho}_{1(2)}=\rho_{1(2)}/\rho_0$ and $\bar{\kappa}_{1(2)}=\kappa_{1(2)}/\kappa_0$, where $\kappa_{1(2)}$ is the compressibility of the top (bottom) particle. The resulting relative material and size parameters for the downward Kerker effect are radii $r_1=0.0282\lambda$, $\bar{\rho}_1=0.837$, $\bar{\kappa}_1=25$, $r_2=0.038\lambda$, $\bar{\rho}_2=0.66$, and $\bar{\kappa}_2=35$; for the upward Kerker effect, the bottom and top particles are swapped. The results of the full-wave simulations for the radiation patterns are plotted as black dots in Fig.~\ref{fig:kerker_maps}(a), showing excellent agreement with the analytical calculations. It should be mentioned that even better agreement can be achieved by taking into account the relatively small effective quadrupole moment; see the Supplemental Information for details. The Kerker condition shifts with changing interparticle distance, as shown by the dashed line in Fig.~\ref{fig:kerker_maps}(b). For $d\gtrsim0.35\lambda$, the effective quadrupole moment is too strong for the model to be truncated at the moments specified by Eq.~\eqref{eq:eff_moments}; see the Supplemental Information for details.

Finally, we also predict that the regime of transverse unidirectional scattering is accessible using realistic structures, with the dimer consisting of labyrinthine acoustic meta-atoms~\cite{cheng2015}; see Fig.~\ref{fig:metaatom}(a). We perform full-wave numerical simulations in COMSOL Multiphysics. Although the homogeneous high-index acoustic materials assumed in the infinitely long cylinder calculations above are difficult to realize in airborne acoustics, the same monopolar response can be implemented using labyrinthine meta-atoms, which can be fabricated by conventional polymer 3D printing. These structures emulate high-index Mie scatterers in air by elongating the internal propagation path and supporting strong subwavelength resonances; their geometry can therefore be optimized to tune the monopole phase while keeping the dipole response weak~\cite{timankova2026experimental}. At a separation $d=0.2\lambda$, which we choose to accommodate practical meta-atom sizes, the target Mie angles in a model of purely monopolar particles are $\theta_{m1}=-0.676$ and $\theta_{m2}=0.271$, i.e., close to resonance for both particles. To optimize the geometry for the required monopolar response, we calculate the phase $\arg\{-\gamma_0\}$ and amplitude $|\gamma_0|$ spectra of the monopolar scattering coefficient; see Fig.~\ref{fig:metaatom} and the SI for more details. For a meta-atom with relatively small Willis~\cite{quan2018maximum} and higher-order~\cite{Smagin2024,yu2025bianisotropic} couplings, under an $\hat{\mathbf{x}}$-oriented incident plane-wave with unit amplitude, the coefficient can be approximately treated as the Mie coefficient $\gamma_0\approx a_0=-\cos(\theta_n)e^{i\theta_n}$. As a result of the geometry optimization, the top and bottom meta-atoms, which have identical geometries but different sizes, have the necessary monopole responses at frequency $f=332.7$~Hz, and the designed dimer exhibits the expected transverse unidirectional scattering; see Fig.~\ref{fig:metaatom}(a). The shape of the numerically calculated radiation pattern does not perfectly correspond to the downward cardioid because each realized labyrinthine particle retains a small residual dipolar response, yet it remains sufficiently close to the analytical result expected from the model with purely monopolar particles, Eq.~\eqref{eq:eff_moments}. The proposed configuration is directly compatible with existing experimental platforms based on parallel-plate waveguides that simulate quasi-2D space~\cite{timankova2026experimental,jordaan2018}. For the realization considered here, the operating frequency is approximately $333$~Hz, and the constituent meta-atoms have outer radii of approximately $3$~cm and a channel width of $0.2$~cm, which is accessible with 3D-printing technology. The proposed geometry can also be shifted to other frequency ranges by scaling the meta-atoms. The transverse directional-scattering pattern could be characterized by scanning the scattered pressure field with a microphone array along a circular contour surrounding the dimer. However, dissipative losses can shift the optimal Mie-angle conditions required for transverse directional scattering; see the SI. Therefore, realistic experimental realizations may require accounting for thermoviscous losses when optimizing the meta-atom geometry.

In conclusion, we have shown that in a single non-absorbing rotationally symmetric acoustic particle, i.e., a sphere or infinitely long cylinder, the directional-scattering conditions are reached only in the limit of vanishing total scattering. In contrast, we predicted that a coupled acoustic dimer overcomes this fundamental restriction and enables a pronounced directional response in the strong-scattering regime. Most importantly, we have used acoustic dimers to theoretically and numerically predict transverse Kerker scattering in acoustics for the first time, in particular unidirectional side scattering. These results and the developed analytical models help establish coupled acoustic meta-atoms as a promising platform for controlling sound with subwavelength structures and open the way toward new acoustic devices and metasurfaces for beam steering, directional routing, and advanced wavefront manipulation.

The published version and Supplemental Information for this work is available at  \url{https://doi.org/10.1016/j.photonics.2026.101594}.

\begin{acknowledgments}
The authors are thankful to Vladimir Igoshin for fruitful discussions. The work is funded by Russian Science Foundation grant № 25-79-31027, https://rscf.ru/project/25-79-31027/. This work was supported by the Ministry of Science and Higher Education of the Russian Federation (Project FSER-2025–0012). The work of PSP was financially supported by the ITMO Fellowship Program. 
\end{acknowledgments}

\bibliography{apssamp}

@article{liu2018,
  title={Generalized Kerker effects in nanophotonics and meta-optics},
  author={Liu, Wei and Kivshar, Yuri S},
  journal={Optics express},
  volume={26},
  number={10},
  pages={13085--13105},
  year={2018},
  publisher={Optical Society of America}
}

@article{kerker1983,
  title={Electromagnetic scattering by magnetic spheres},
  author={Kerker, Milton and Wang, D-S and Giles, CL},
  journal={Journal of the Optical Society of America},
  volume={73},
  number={6},
  pages={765--767},
  year={1983},
  publisher={Optical Society of America}
}

@article{wei2020,
  title={Far-field and near-field directionality in acoustic scattering},
  author={Wei, Lei and Rodr{\'\i}guez-Fortu{\~n}o, Francisco J},
  journal={New Journal of Physics},
  volume={22},
  number={8},
  pages={083016},
  year={2020},
  publisher={IOP Publishing}
}

@article{wu2021,
  title={Acoustic generalized Kerker effect},
  author={Wu, Hong-Wei and Li, Wei and Cheng, Shu-Ling and Yin, Yun-Qiao and Sheng, Zong-Qiang},
  journal={Applied Physics Express},
  volume={14},
  number={9},
  pages={095501},
  year={2021},
  publisher={IOP Publishing}
}

@article{Smagin2024,
  title={Acoustic lateral recoil force and stable lift of anisotropic particles},
  author={Smagin, Mikhail and Toftul, Ivan and Bliokh, Konstantin Y and Petrov, Mihail},
  journal={Physical Review Applied},
  volume={22},
  number={6},
  pages={064041},
  year={2024},
  publisher={APS}
}

@article{toftul2019,
  title={Acoustic radiation force and torque on small particles as measures of the canonical momentum and spin densities},
  author={Toftul, ID and Bliokh, KY and Petrov, Mihail I and Nori, Franco},
  journal={Physical Review Letters},
  volume={123},
  number={18},
  pages={183901},
  year={2019},
  publisher={APS}
}

@article{Sun2020,
  title={Broadband acoustic ventilation barriers},
  author={Sun, Man and Fang, Xinsheng and Mao, Dongxing and Wang, Xu and Li, Yong},
  journal={Physical Review Applied},
  volume={13},
  number={4},
  pages={044028},
  year={2020},
  publisher={APS}
}

@article{cheng2015,
  title={Ultra-sparse metasurface for high reflection of low-frequency sound based on artificial Mie resonances},
  author={Cheng, Y and Zhou, C and Yuan, BG and Wu, DJ and Wei, Q and Liu, XJ},
  journal={Nature Materials},
  volume={14},
  number={10},
  pages={1013--1019},
  year={2015},
  publisher={Nature Publishing Group UK London}
}

@book{blackstock2000,
  author    = {D. T. Blackstock},
  title     = {Fundamentals of Physical Acoustics},
  publisher = {John Wiley \& Sons},
  address   = {New York},
  year      = {2000},
}

@article{jordaan2018,
  title={Measuring monopole and dipole polarizability of acoustic meta-atoms},
  author={Jordaan, Joshua and Punzet, Stefan and Melnikov, Anton and Sanches, Alexandre and Oberst, Sebastian and Marburg, Steffen and Powell, David A},
  journal={Applied Physics Letters},
  volume={113},
  number={22},
  year={2018},
  publisher={AIP Publishing}
}

@article{toftul2025acoustic,
  title={Acoustic angular sorting of resonant subwavelength particles},
  author={Toftul, Ivan and Kivshar, Yuri S and Lapine, Mikhail},
  journal={Physical Review Applied},
  volume={24},
  number={3},
  pages={034019},
  year={2025},
  publisher={APS}
}

@article{yu2022broadband,
  title={Broadband unidirectional transverse light scattering in a V-shaped silicon nanoantenna},
  author={Yu, Yang and Liu, Jinze and Yu, Yidu and Qiao, Dayong and Li, Yongqian and Salas-Montiel, Rafael},
  journal={Optics Express},
  volume={30},
  number={5},
  pages={7918--7927},
  year={2022},
  publisher={Optica Publishing Group}
}

@article{rahimzadegan2020minimalist,
  title={Minimalist Mie coefficient model},
  author={Rahimzadegan, Aso and Alaee, Rasoul and Rockstuhl, Carsten and Boyd, Robert W},
  journal={Optics Express},
  volume={28},
  number={11},
  pages={16511--16525},
  year={2020},
  publisher={Optical Society of America}
}

@article{yosioka,
	Author = {Yosioka, K. and Kawasima, Y.},
	Journal = {Acta Acust. United Acust.},
	Number = {3},
	Pages = {167--173},
	Title = {{Acoustic radiation pressure on a compressible sphere}},
	Url = {https://www.ingentaconnect.com/content/dav/aaua/1955/00000005/00000003/art00004},
	Volume = {5},
	Year = {1955}}

@article{timankova2026experimental,
  title={Experimental Investigation of Acoustic Kerker Effect in Labyrinthine Resonators},
  author={Timankova, IA and Smagin, MV and Kuzmin, MV and Lutovinov, AI and Bogdanov, AA and Li, Yong and Petrov, MI},
  journal={JETP Letters},
  pages={1--7},
  year={2026},
  publisher={Springer}
}

@article{zheng2023annular,
  title={Annular and unidirectional transverse scattering with high directivity based on magnetoelectric coupling},
  author={Zheng, Kaihao and Li, Wenjia and Sun, Botian and Wang, Yehan and Guan, Chunying and Liu, Jianlong and Shi, Jinhui},
  journal={Optics Express},
  volume={31},
  number={9},
  pages={14037--14047},
  year={2023},
  publisher={Optica Publishing Group}
}

@article{bag2018transverse,
  title={Transverse kerker scattering for angstrom localization of nanoparticles},
  author={Bag, Ankan and Neugebauer, Martin and Wo{\'z}niak, Pawe{\l} and Leuchs, Gerd and Banzer, Peter},
  journal={Physical Review Letters},
  volume={121},
  number={19},
  pages={193902},
  year={2018},
  publisher={APS}
}

@article{lee2018simultaneously,
  title={Simultaneously nearly zero forward and nearly zero backward scattering objects},
  author={Lee, Jeng Yi and Miroshnichenko, Andrey E and Lee, Ray-Kuang},
  journal={Optics Express},
  volume={26},
  number={23},
  pages={30393--30399},
  year={2018},
  publisher={Optical Society of America}
}

@article{shamkhi2019transverse,
  title={Transverse scattering and generalized Kerker effects in all-dielectric Mie-resonant metaoptics},
  author={Shamkhi, Hadi K and Baryshnikova, Kseniia V and Sayanskiy, Andrey and Kapitanova, Polina and Terekhov, Pavel D and Belov, Pavel and Karabchevsky, Alina and Evlyukhin, Andrey B and Kivshar, Yuri and Shalin, Alexander S},
  journal={Physical Review Letters},
  volume={122},
  number={19},
  pages={193905},
  year={2019},
  publisher={APS}
}

@article{bukharin2022transverse,
  title={Transverse Kerker effect in all-dielectric spheroidal particles},
  author={Bukharin, Mikhail M and Pecherkin, Vladimir Ya and Ospanova, Anar K and Il’in, Vladimir B and Vasilyak, Leonid M and Basharin, Alexey A and Luk ‘yanchuk, Boris},
  journal={Scientific Reports},
  volume={12},
  number={1},
  pages={7997},
  year={2022},
  publisher={Nature Publishing Group UK London}
}

@article{liu2019lattice,
  title={Lattice invisibility effect based on transverse Kerker scattering in 1D metalattices},
  author={Liu, MQ and Zhao, CY},
  journal={Journal of Physics D: Applied Physics},
  volume={52},
  number={49},
  pages={495107},
  year={2019},
  publisher={IOP Publishing}
}

@article{du2024manipulation,
  title={Manipulation of unidirectional side scattering of light in transition metal dichalcogenide nanoresonators},
  author={Du, Jing and Bao, Wenrui and Zhang, Ruiyang and Shen, Shiyu and Yue, Lin and Ding, Qi and Xie, Peng and Kuang, Xiaoyu and Zhang, Hong and Wang, Wei},
  journal={Physical Review B},
  volume={109},
  number={11},
  pages={115426},
  year={2024},
  publisher={APS}
}

@article{matsumori2023unidirectional,
  title={Unidirectional transverse light scattering in notched silicon nanosphere},
  author={Matsumori, Akira and Sugimoto, Hiroshi and Fujii, Minoru},
  journal={Laser \& Photonics Reviews},
  volume={17},
  number={8},
  pages={2300314},
  year={2023},
  publisher={Wiley Online Library}
}

@article{shamkhi2019transparency,
  title={Transparency and perfect absorption of all-dielectric resonant metasurfaces governed by the transverse Kerker effect},
  author={Shamkhi, Hadi K and Sayanskiy, Andrey and Valero, Adri{\`a} Can{\'o}s and Kupriianov, Anton S and Kapitanova, Polina and Kivshar, Yuri S and Shalin, Alexander S and Tuz, Vladimir R},
  journal={Physical Review Materials},
  volume={3},
  number={8},
  pages={085201},
  year={2019},
  publisher={APS}
}

@article{zhang2021wide,
  title={Wide-angle invisible dielectric metasurface driven by transverse Kerker scattering},
  author={Zhang, Xia and Bradley, A Louise},
  journal={Physical Review B},
  volume={103},
  number={19},
  pages={195419},
  year={2021},
  publisher={APS}
}

@article{nechayev2019huygens,
  title={Huygens' dipole for polarization-controlled nanoscale light routing},
  author={Nechayev, Sergey and Eismann, J{\"o}rg S and Neugebauer, Martin and Wo{\'z}niak, Pawe{\l} and Bag, Ankan and Leuchs, Gerd and Banzer, Peter},
  journal={Physical Review A},
  volume={99},
  number={4},
  pages={041801},
  year={2019},
  publisher={APS}
}

@article{li2016all,
  title={All-dielectric antenna wavelength router with bidirectional scattering of visible light},
  author={Li, Jiaqi and Verellen, Niels and Vercruysse, Dries and Bearda, Twan and Lagae, Liesbet and Van Dorpe, Pol},
  journal={Nano Letters},
  volume={16},
  number={7},
  pages={4396--4403},
  year={2016},
  publisher={ACS Publications}
}

@article{qin2022transverse,
  title={Transverse Kerker effect for dipole sources},
  author={Qin, Feifei and Zhang, Zhanyuan and Zheng, Kanpei and Xu, Yi and Fu, Songnian and Wang, Yuncai and Qin, Yuwen},
  journal={Physical Review Letters},
  volume={128},
  number={19},
  pages={193901},
  year={2022},
  publisher={APS}
}

@article{zhang2024magnetic,
  title={Magnetic transverse unidirectional scattering and longitudinal displacement sensing in silicon nanodimer},
  author={Zhang, Zhaokun and Xu, Jipeng and Liu, Ken and Zhu, Zhihong},
  journal={Optics Express},
  volume={32},
  number={11},
  pages={19279--19293},
  year={2024},
  publisher={Optica Publishing Group}
}

@article{zhang2021nanometric,
  title={Nanometric displacement sensor with a switchable measuring range using a cylindrical vector beam excited silicon nanoantenna},
  author={Zhang, Hanmou and Gao, Kun and Han, Lei and Liu, Sheng and Mei, Ting and Xiao, Fajun and Zhao, Jianlin},
  journal={Optics Express},
  volume={29},
  number={16},
  pages={25109--25117},
  year={2021},
  publisher={Optical Society of America}
}

@article{shang2019unidirectional,
  title={Unidirectional scattering exploited transverse displacement sensor with tunable measuring range},
  author={Shang, Wuyun and Xiao, Fajun and Zhu, Weiren and Han, Lei and Premaratne, Malin and Mei, Ting and Zhao, Jianlin},
  journal={Optics Express},
  volume={27},
  number={4},
  pages={4944--4955},
  year={2019},
  publisher={Optical Society of America}
}

@article{liu2017beam,
  title={Beam steering with dielectric metalattices},
  author={Liu, Wei and Miroshnichenko, Andrey E},
  journal={ACS Photonics},
  volume={5},
  number={5},
  pages={1733--1741},
  year={2017},
  publisher={ACS Publications}
}

@article{hancu2014multipolar,
  title={Multipolar interference for directed light emission},
  author={Hancu, Ion M and Curto, Alberto G and Castro-L{\'o}pez, Marta and Kuttge, Martin and van Hulst, Niek F},
  journal={Nano Letters},
  volume={14},
  number={1},
  pages={166--171},
  year={2014},
  publisher={ACS Publications}
}

@article{barreda2017electromagnetic,
  title={Electromagnetic polarization-controlled perfect switching effect with high-refractive-index dimers and the beam-splitter configuration},
  author={Barreda, Angela I and Saleh, Hassan and Litman, Amelie and Gonz{\'a}lez, Francisco and Geffrin, Jean-Michel and Moreno, Fernando},
  journal={Nature Communications},
  volume={8},
  number={1},
  pages={13910},
  year={2017},
  publisher={Nature Publishing Group UK London}
}

@article{rutckaia2017quantum,
  title={Quantum dot emission driven by Mie resonances in silicon nanostructures},
  author={Rutckaia, Viktoriia and Heyroth, Frank and Novikov, Alexey and Shaleev, Mikhail and Petrov, Mihail and Schilling, Joerg},
  journal={Nano Letters},
  volume={17},
  number={11},
  pages={6886--6892},
  year={2017},
  publisher={ACS Publications}
}

@book{benesty2012study,
  title={Study and design of differential microphone arrays},
  author={Benesty, Jacob and Jingdong, Chen},
  volume={6},
  year={2012},
  publisher={Springer Science \& Business Media}
}

@inproceedings{elko1995simple,
  title={A simple adaptive first-order differential microphone},
  author={Elko, Gary W and Pong, Anh-Tho Nguyen},
  booktitle={Proceedings of 1995 Workshop on Applications of Signal Processing to Audio and Accoustics},
  pages={169--172},
  year={1995},
  organization={IEEE}
}

@book{kim2013sound,
  title={Sound visualization and manipulation},
  author={Kim, Yang-Hann and Choi, Jung-Woo},
  year={2013},
  publisher={John Wiley \& Sons}
}

@article{olson1973gradient,
  title={Gradient loudspeakers},
  author={Olson, Harry F},
  journal={Journal of the Audio Engineering Society},
  volume={21},
  pages={86–93},
  year={1973}
}

@article{quan2018maximum,
  title={Maximum Willis coupling in acoustic scatterers},
  author={Quan, Li and Ra’di, Younes and Sounas, Dimitrios L and Al{\`u}, Andrea},
  journal={Physical Review Letters},
  volume={120},
  number={25},
  pages={254301},
  year={2018},
  publisher={APS}
}

@article{boya1994optical,
  title={Optical theorem in N dimensions},
  author={Boya, Luis J and Murray, Robert},
  journal={Physical Review A},
  volume={50},
  number={5},
  pages={4397},
  year={1994},
  publisher={APS}
}

@article{yu2025bianisotropic,
  title={Bianisotropic response of asymmetrical acoustic scatterers, including the quadrupole couplings},
  author={Yu, Gaokun and Shi, Zijian},
  journal={Physical Review Applied},
  volume={24},
  number={3},
  pages={034059},
  year={2025},
  publisher={APS}
}

@article{alu2010does,
  title={How does zero forward-scattering in magnetodielectric nanoparticles comply with the optical theorem?},
  author={Alu, Andrea and Engheta, Nader},
  journal={Journal of Nanophotonics},
  volume={4},
  number={1},
  pages={041590},
  year={2010},
  publisher={SPIE}
}

@article{toftul2026radiation,
  title={Radiation forces and torques in optics and acoustics},
  author={Toftul, Ivan and Golat, Sebastian and Rodr{\'\i}guez-Fortu{\~n}o, Francisco J and Nori, Franco and Kivshar, Yuri and Bliokh, Konstantin Y},
  journal={Reviews of Modern Physics},
  volume={98},
  number={2},
  pages={025002},
  year={2026},
  publisher={APS}
}

@article{ustimenko2026lattice,
  title={Lattice-induced sound trapping in biperiodic metasurfaces of acoustic resonators},
  author={Ustimenko, Nikita and Evlyukhin, Andrey B and Kyrimi, Vicky and Kildishev, Alexander V and Rockstuhl, Carsten},
  journal={Physical Review Research},
  volume={8},
  number={1},
  pages={013074},
  year={2026},
  publisher={APS}
}

\appendix
\section{Proof of the Non-Existence of the Kerker Effect in a Non-Absorbing Rotationally Symmetric Particle}
\label{appendix:proof}

 In general, for a scalar Green's function $G(\omega,\bm r)$, the pressure fields radiated by the monopole $M$ and dipole $\bm D$ moments in the $\eu^{-\,\iu\omega t}$ convention can be defined as:
\begin{align}
p_M(\bm r) &= \rho \omega^2 M\,G(\omega,\bm r), \label{eqS:pM_general}\\
p_D(\bm r) &= -\,\iu \rho c k\,\big(\nabla G(\omega,\bm r)\cdot \bm D\big).
\label{eqS:pD_general}
\end{align}
For 2D free space,
\begin{equation}
G(\bm r)=\frac{\iu}{4}H_0^{(1)}(kr),
\end{equation}
so that
\begin{equation}
\nabla G(\bm r)= -\frac{\iu k}{4}H_1^{(1)}(kr)\,\hat{\bm n},
\qquad
\hat{\bm n}=(\cos\phi,\sin\phi).
\label{eqS:gradG}
\end{equation}
Substituting \eqref{eqS:gradG} into \eqref{eqS:pM_general} and \eqref{eqS:pD_general}, the monopole--dipole pressure fields become
\begin{align}
p_M(r) &= \frac{\iu}{4}\,\rho c^2 k^2\, M\, H_0^{(1)}(kr), \label{eqS:pM}\\
p_D(r) &= -\frac{1}{4}\,\rho c\, k^2\, H_1^{(1)}(kr)\,(\hat{\bm n}\cdot \bm D).
\label{eqS:pD}
\end{align}
Using $H_1^{(1)}(kr)\simeq -\iu H_0^{(1)}(kr)$ for $kr\gg1$, the far-field scattered pressure reads
\begin{equation}
p_{\rm sc}(r,\phi)\simeq \frac{\iu}{4}\,\rho c^2 k^2\, H_0^{(1)}(kr)\Big(M+\frac{1}{c}\hat{\bm n}\cdot\bm D\Big).
\label{eqS:farfield}
\end{equation}
For plane-wave incidence along $+x$, a rotationally symmetric particle induces $\bm D=D_x\hat{\bm x}$, so that $\hat{\bm n}\cdot\bm D=D_x\cos\phi$. Evaluating Eq.~\eqref{eqS:farfield} at $\phi=0$ (forward) and $\phi=\pi$ (backward) gives
\begin{equation}
p_{\rm fwd}\propto M+\frac{D_x}{c},\qquad
p_{\rm back}\propto M-\frac{D_x}{c}.
\label{eqS:pfwdback}
\end{equation}
We define polarizabilities (single particle)
\begin{equation}
M=\alpha_m\ \kappa \,p_{\rm loc},\qquad \bm D=\alpha_d\,\bm v_{\rm loc},
\label{eqS:pol_def}
\end{equation}
where, for a rotationally symmetric particle, $\alpha_d$ is scalar and $\bm D\parallel\bm v_{\rm loc}$.

The scattered and incident fields may be expanded into cylindrical harmonics:
\begin{align}
p_{\rm inc}(r,\phi)
&=
p_0\sum_{n=-\infty}^{\infty}
\beta_n J_n(kr)\eu^{\iu n\phi},
\\
p_{\rm sc}(r,\phi)
&=
p_0\sum_{n=-\infty}^{\infty}
\gamma_n H_n^{(1)}(kr)\eu^{\iu n\phi}
\nonumber\\
&=
p_0\sum_{n=-\infty}^{\infty}
\beta_n a_n H_n^{(1)}(kr)\eu^{\iu n\phi},
\label{eqS:mieexp}
\end{align}
so that the partial-wave scattering coefficient is $\gamma_n=a_n\beta_n$. For an infinitely long circular cylinder of radius $a$ (2D scattering), the Mie coefficients $a_n$ in Eq.~\eqref{eqS:mieexp} follow from enforcing continuity of pressure and normal particle velocity at $r=a$. For a homogeneous fluid cylinder with interior wavenumber $k_1$ and density $\rho_1$ embedded in a background $(k,\rho)$, the coefficients take the form~\cite{yosioka}:
\begin{equation}
a_n
=
\frac{\frac{\rho_1 k}{\rho\,k_1}J_n'(k_1 a)\,J_n(k a)-J_n(k_1 a)\,J_n'(k a)}
     {H_n^{(1)\prime}(k a)\,J_n(k_1 a)-\frac{\rho_1 k}{\rho\,k_1}J_n'(k_1 a)\,H_n^{(1)}(k a)},
\label{eqS:cyl_mie}
\end{equation}
where $J_n$ and $H_n^{(1)}$ are Bessel and outgoing Hankel functions, respectively, and primes denote derivatives
with respect to the argument.
%In the rigid (sound-hard) limit this reduces to
%\begin{equation}
%a_n^{\mathrm{rigid}}=-\,\frac{J_n'(k a)}{H_n^{(1)\prime}(k a)}.
%\label{eqS:cyl_mie_rigid}
%\end{equation}
The corresponding 3D expressions for spherical scatterers are obtained by replacing cylindrical Bessel/Hankel
functions by the spherical ones. In the monopole-dipole regime only $a_0$ and $a_1$ are retained.
In two dimension the relation between polarizabilities and Mie coefficients is:
\begin{equation}
\alpha_m=\frac{4a_0}{\iu k^2},\qquad \alpha_d=\frac{8a_1}{\iu k^2}.
\label{eqS:alpha_mie}
\end{equation}

For any non-absorbing rotationally symmetric particle, each Mie coefficient can be parametrized by a single parameter $\theta_n$~\cite{rahimzadegan2020minimalist,toftul2025acoustic}:
\begin{equation}
a_n=-\cos(\theta_n)\,\eu^{\iu\theta_n},\qquad \theta_n\in\Big(-\frac{\pi}{2},\frac{\pi}{2}\Big),
\label{eqS:mie_angle}
\end{equation}
and we denote $\theta_m:=\theta_0$ (monopole angle) and $\theta_d:=\theta_1$ (dipole angle).

%\subsection{Non-existence of nontrivial 1st and 2nd Kerker conditions}

Using \eqref{eqS:pfwdback}--\eqref{eqS:mie_angle} (and the fact that the dipolar channel in 2D comes from the pair $n=\pm1$),
the backward and forward amplitudes can be written (up to an overall common prefactor/sign) as
\begin{align}
p_{\rm back}(\theta_m,\theta_d) &\propto \cos(\theta_m)\eu^{\iu\theta_m}-2\cos(\theta_d)\eu^{\iu\theta_d},\label{eqS:pback_angles}\\
p_{\rm fwd}(\theta_m,\theta_d)  &\propto \cos(\theta_m)\eu^{\iu\theta_m}+2\cos(\theta_d)\eu^{\iu\theta_d}. \label{eqS:pfwd_angles}
\end{align}

\paragraph*{(i) First Kerker condition: $p_{\rm back}=0$.}
Setting \eqref{eqS:pback_angles} to zero gives
\begin{equation}
\cos(\theta_m)\eu^{\iu\theta_m}=2\cos(\theta_d)\eu^{\iu\theta_d}.
\label{eqS:kerker1_eq}
\end{equation}
Taking absolute values yields $\cos(\theta_m)=2\cos(\theta_d)$.
If both cosines are strictly positive, dividing \eqref{eqS:kerker1_eq} by the positive reals gives
$\eu^{\iu\theta_m}=\eu^{\iu\theta_d}$ and hence $\theta_m=\theta_d$ within $(-\pi/2,\pi/2)$, which contradicts
$\cos(\theta_m)=2\cos(\theta_m)$ unless $\cos(\theta_m)=0$.
Therefore at least one cosine must vanish; from \eqref{eqS:kerker1_eq} it follows that
\begin{equation}
p_{\rm back}=0 \quad\Longleftrightarrow\quad \cos(\theta_m)=0 \ \ \text{and}\ \ \cos(\theta_d)=0,
\label{eqS:kerker1_only_trivial}
\end{equation}
i.e. the only solutions correspond to $a_0=a_1=0$ (trivial zero-scattering limit).

\setcounter{figure}{0}
\renewcommand{\thefigure}{B\arabic{figure}}

\begin{figure*}[t!]
  \centering
  \includegraphics[width= 1.0\linewidth]{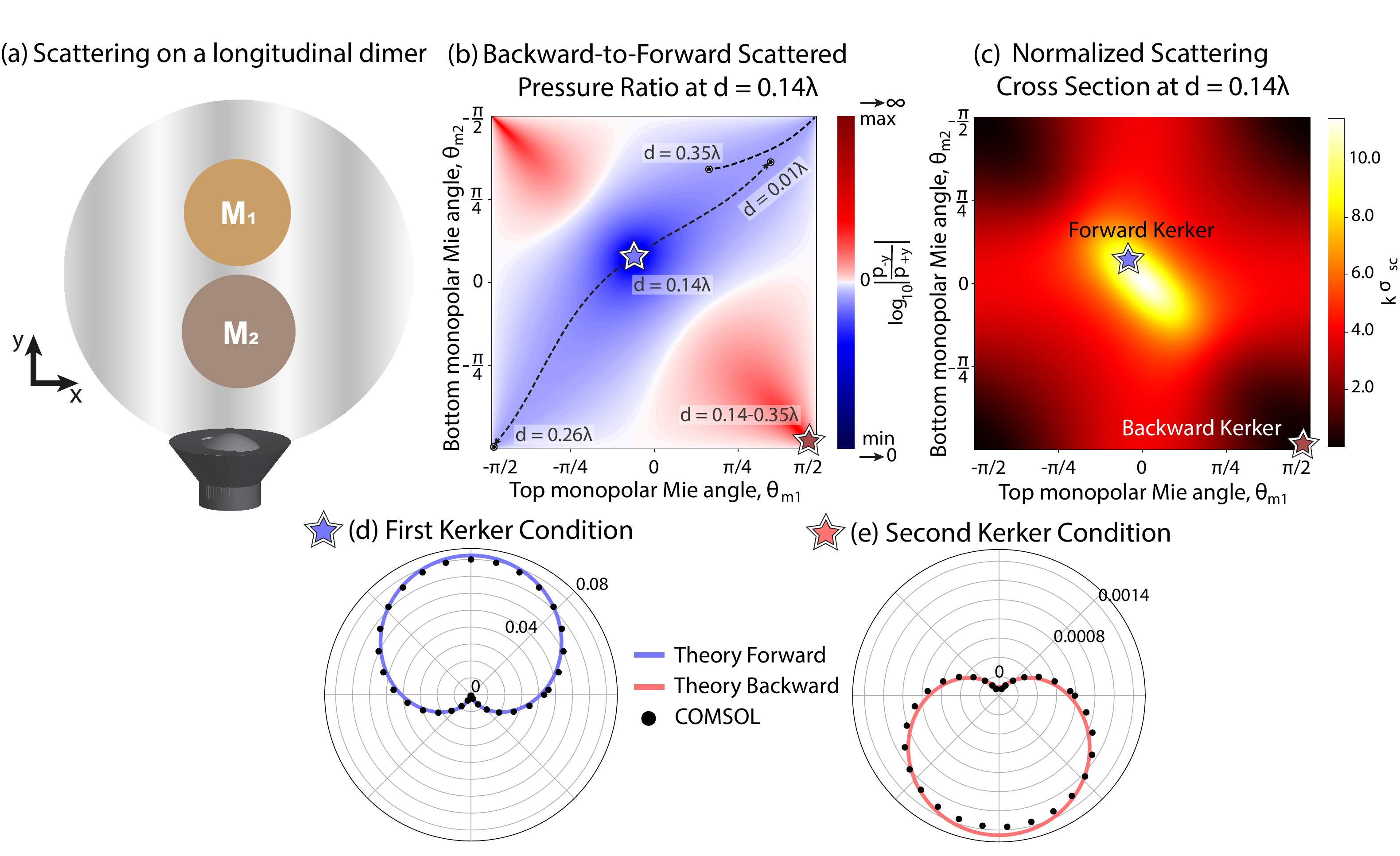}
  \caption{\textbf{Longitudinal Kerker Effect in Acoustic Dimers.}
(a) Acoustic dimer excited by the incident plane-wave along the dimer axis, $\mathbf{k}\parallel\hat{\mathbf{y}}$.
(b) Backward-to-forward logarithmic scattered-pressure ratio of a dimer with interparticle distance $d=0.14\lambda$ in the parameter space of the top and bottom monopolar Mie angles under longitudinal plane-wave excitation.
(c) Normalized scattering cross section $k\,\sigma_{\mathrm{sc}}$ of a monopolar dimer with interparticle distance $d=0.14\lambda$ in the parameter space of monopole Mie angles under longitudinal plane-wave excitation. The stars correspond to the forward and backward Kerker conditions.
(d,e) Analytically and numerically calculated radiation patterns for a dimer with interparticle distance $d=0.14\lambda$ at the forward and backward Kerker conditions.
 }
  \label{fig:longitudinal_kerker}
\end{figure*}

\paragraph*{(ii) Second Kerker condition: $p_{\rm fwd}=0$.}
Similarly, $p_{\rm fwd}=0$ implies
\begin{equation}
\cos(\theta_m)\eu^{\iu\theta_m}=-2\cos(\theta_d)\eu^{\iu\theta_d}.
\label{eqS:kerker2_eq}
\end{equation}
Taking absolute values again yields $\cos(\theta_m)=2\cos(\theta_d)$.
If both cosines are strictly positive, dividing \eqref{eqS:kerker2_eq} gives
$\eu^{\iu(\theta_m-\theta_d)}=-1$, i.e. $\theta_m-\theta_d=\pi \ (\mathrm{mod}\ 2\pi)$,
which has \emph{no} solution for $\theta_m,\theta_d\in(-\pi/2,\pi/2)$.
Hence, as above, the only possible zeros are the trivial corner points with $a_0=a_1=0$. We conclude that perfect forward/backward cancellation for a non-absorbing rotationally symmetric particle in the monopole--dipole approximation is only possible in the trivial vanishing-channel limit.

\textbf{This result is also true in three dimensions.} In order to obtain it, one needs to use the three-dimensional connection between polarizabilities and Mie coefficients~\cite{toftul2019}:
\begin{equation}
\alpha_m=\frac{4\pi a_0}{\iu k^3},
\qquad
\alpha_d=\frac{12\pi a_1}{\iu k^3},
\label{eqS:alpha_mie_3D}
\end{equation}
and the rest of the derivation is similar. True full scattering cancellation resulting from interference of the two lowest channels for the simplest acoustic particles is therefore possible only in the one-dimensional case~\cite{Sun2020}.

\section{Longitudinal Kerker Effect in Dimers}
\label{appendix:longitudinal}

Here we consider the case of longitudinal incidence, when the plane-wave propagates along the dimer axis, $\mathbf{k}\parallel\hat{\mathbf y}$. In this geometry, the directional response is measured in the forward and backward directions, so that the dimer realizes the conventional Kerker effect. Unlike a single non-absorbing rotationally symmetric particle, however, the coupled dimer can achieve directional cancellation while retaining strong overall scattering. Analogously to the transverse Kerker effect, we limit our attention to the monopolar-particle model.

We take the incident plane wave convention with incidence angle $\theta=\pi/2$, see Fig.\ref{fig:longitudinal_kerker}(a), so that:
\begin{equation}
p^{\mathrm{inc}}(\mathbf r)=p_0 e^{i k \,y}
\end{equation}
The incident fields at particle positions are then:
\begin{equation}
p^{\mathrm{inc}}(\mathbf r_{1,2}) = p_0\,e^{\pm i k d/2},\qquad
\mathbf v^{\mathrm{inc}}(\mathbf r_{1,2}) = \frac{p_0}{\rho c}\,e^{\pm i k d/2}\,\hat{\mathbf y},
\label{eq:drive_longitudinal}
\end{equation}
so that only the $v_y$ component is directly excited, while the two particles are driven with opposite phase factors set by $kd$. 

To quantify the directionality, we plot explicitly the ratio of backward to forward scattered pressure:
\begin{equation}
\frac{p_{-y}}{p_{+y}} = \frac{J_0\!\left(\tfrac{kd}{2}\right)(M_1+M_2)-2iJ_1\!\left(\tfrac{kd}{2}\right)(M_2-M_1)}{J_0\!\left(\tfrac{kd}{2}\right)(M_1+M_2)+2iJ_1\!\left(\tfrac{kd}{2}\right)(M_2-M_1)}.
\label{eq:ratio_longitudinal}
\end{equation}
With this convention, a minimum of $\left|p_{-y}/p_{+y}\right|$ corresponds to suppression of the backward-scattered field $p_{-y}$ and therefore to the first Kerker condition, or forward Kerker effect. Conversely, a maximum corresponds to suppression of the forward-scattered field $p_{+y}$, i.e. to the second Kerker condition or backward Kerker effect (anti-Kerker effect).

Figure~\ref{fig:longitudinal_kerker}(b) shows $\left|p_{-y}/p_{+y}\right|$ in the plane of monopolar Mie angles $(\theta_{m1},\theta_{m2})$ for $d=0.14\lambda$. The blue star marks the minimum associated with the first Kerker condition, while the red star marks the maximum associated with the second Kerker condition. The dashed trajectory indicates how these extrema evolve with changing interparticle distance.

The corresponding normalized scattering cross section is shown in Fig.~\ref{fig:longitudinal_kerker}(c). In contrast to the single-particle case, the minimum of $\left|p_{-y}/p_{+y}\right|$ lies in a bright high-scattering region, showing that the dimer can suppress backward scattering without approaching the weak-scattering limit. This represents the qualitative difference from the single rotationally symmetric particle discussed in the main text. The second Kerker point, by comparison, lies much closer to the weak-scattering boundary, or a corner in the Mie-angle space, so that forward suppression is accompanied by a substantially smaller total response, which is a direct consequence of the optical theorem~\cite{boya1994optical,alu2010does}.

The radiation patterns in Fig.~\ref{fig:longitudinal_kerker}(d,e) correspond to the starred points in Fig.~\ref{fig:longitudinal_kerker}(b). Analytical calculations are performed in the monopolar-particle model truncated at the dipole order, and COMSOL calculations are performed on infinitely long 2D cylinders. For the forward Kerker condition, the target angles are $\theta_{m1}=-0.198$ and $\theta_{m2}=0.229$, and the particle parameters are radius $r_1=0.0306\lambda$, relative density $\bar{\rho}_1=0.78$, relative compressibility $\bar{\kappa}_1=25$, $r_2=0.031\lambda$, $\bar{\rho}_2=0.72$, and $\bar{\kappa}_2=35$. For the backward Kerker condition, the target angles are $\theta_{m1}=\pi/2.0361$ and $\theta_{m2}=-\pi/2.0361$, and the particle parameters are $r_1=0.047\lambda$, $\bar{\rho}_1=25$, $\bar{\kappa}_1=7.32$, $r_2=0.0227\lambda$, $\bar{\rho}_2=22.3$, and $\bar{\kappa}_2=35$. Here, the material parameters $\bar{\rho}_{1,2}=\rho_{1,2}/\rho_0$ and $\bar{\kappa}_{1,2}=\kappa_{1,2}/\kappa_0$ are given relative to the host medium.
%
% ****** End of file apssamp.tex ******

\end{document}